# Dermatologist-like explainable AI enhances trust and confidence in diagnosing melanoma

Tirtha Chanda[a*], Katja Hauser[a*], Sarah Hobelsberger[b*], Tabea-Clara Bucher[a], Carina Nogueira Garcia[a], Christoph Wies[a], Harald Kittler[c], Philipp Tschandl[c], Cristian Navarrete-Dechent[d], Sebastian Podlipnik[e], Emmanouil Chousakos[f], Iva Crnaric[g], Jovana Majstorovich[h], Linda Alhajwan[i], Tanya Foreman[j], Sandra Peternel[k], Sergei Sarap[l], İrem Özdemir[m], Raymond L. Barnhill[n], Mar Llamas Velasco[o], Gabriela Poch[p], Sören Korsing[p], Wiebke Sondermann[q], Frank Friedrich Gellrich[r], Markus V. Heppt[s], Michael Erdmann[s], Sebastian Haferkamp[t], Konstantin Drexler[t], Matthias Goebeler[u], Bastian Schilling[u], Jochen S. Utikal[v], Kamran Ghoreschi[p], Stefan Fröhling[w], Eva Krieghoff-Henning[a], Titus J. Brinker[†a]

*These authors contributed equally to this work.

a. Digital Biomarkers for Oncology Group, German Cancer Research Center (DKFZ), Heidelberg, Germany
b. Department of Dermatology, University Hospital, Technical University Dresden, Dresden, Germany
c. Department of Dermatology, Medical University of Vienna, Austria
d. Department of Dermatology, Escuela de Medicina, Pontificia Universidad Católica de Chile, Santiago, Chile
e. Dermatology Department, Hospital Clínic of Barcelona, University of Barcelona, IDIBAPS, Barcelona, Spain
f. 1st Department of Pathology, Medical School, National & Kapodistrian University of Athens, Athens, Greece
g. Department of Dermatovenereology, Sestre milosrdnice University Hospital Center, Zagreb, Croatia
h. Derma Style, Dermatovenerology clinic, Belgrade, Serbia
i. Department of Dermatology, Dubai London Clinic, Dubai, United Arab Emirates
j. West Dermatology, Newport Beach, California, USA
k. Department of Dermatovenereology, Clinical Hospital Center Rijeka, Faculty of Medicine, University of Rijeka, Rijeka, Croatia
l. LaserMed, Tallinn, Estonia
m. Department of Dermatology, Faculty of Medicine, Gazi University, Ankara, Turkey
n. Department of Translational Research, Institut Curie, Unit of Formation and Research of Medicine University of Paris, Paris, France
o. Universidad Autónoma de Madrid, Madrid, Spain
p. Charité - Universitätsmedizin Berlin, corporate member of Freie Universität Berlin and Humboldt-Universität zu Berlin, Department of Dermatology, Venereology and Allergology, Berlin, Germany
q. Department of Dermatology, University Hospital Essen, University Duisburg-Essen, Essen, Germany
r. Department of Dermatology and University Hospital Carl Gustav Carus, Technische Universität Dresden, Dresden, Germany
s. Department of Dermatology, Uniklinikum Erlangen, Friedrich-Alexander-Universität Erlangen-Nürnberg, Erlangen, Germany
t. Department of Dermatology, University Hospital Regensburg, Regensburg, Germany
u. Department of Dermatology, Venereology and Allergology, University Hospital Würzburg, Würzburg, Germany

v. Department of Dermatology, Venereology and Allergology, University Medical Center Mannheim, Ruprecht-Karl University of Heidelberg, Mannheim, Germany

w. Division of Translational Medical Oncology, National Center for Tumor Diseases (NCT) Heidelberg and German Cancer Research Center (DKFZ), Heidelberg, Germany

† Correspondence to:

Dr. med. Titus J. Brinker, MD

Digital Biomarkers for Oncology Group, National Center for Tumor Diseases (NCT), German Cancer Research Center (DKFZ), Im Neuenheimer Feld 280, 69120 Heidelberg, Germany

Tel.: +496221 3219304; Email: titus.brinker@dkfz.de

**Conflicts of interest (including 3 years before submission, i.e. 2019-2023):**

PT reports grants from Lilly, consulting fees from Silverchair, lecture honoraria from Lilly, FotoFinder and Novartis, outside of the present publication.

TJB owns a company that develops mobile apps (Smart Health Heidelberg GmbH, Heidelberg, Germany), outside of the scope of the submitted work.

WS received travel support for participation in congresses and / or (speaker) honoraria as well as research grants from medi GmbH Bayreuth, Abbvie, Almirall, Amgen, Bristol-Myers Squibb, Celgene, GSK, Janssen, LEO Pharma, Lilly, MSD, Novartis, Pfizer, Roche, Sanofi Genzyme, and UCB outside of the present publication.

MLV received travel support for participation in congresses and / or (speaker) honoraria as well as research grants from Abbvie, Almirall, Amgen, Bristol-Myers Squibb, Celgene, Janssen, Kyowa Kirin, LEO Pharma, Lilly, MSD, Novartis, Pfizer, Roche, Sanofi Genzyme, and UCB outside of the present publication.

BS is on the advisory board or has received honoraria from Immunocore, Almirall, Pfizer, Sanofi, Novartis, Roche, BMS and MSD, research funding from Novartis and Pierre Fabre Pharmaceuticals, and travel support from Novartis, Roche, Bristol-Myers Squibb and Pierre Fabre Pharma, outside the submitted work.

SH is on the advisory board or has received honoraria from Novartis, Pierre Fabre, BMS and MSD outside the submitted work.

KD has received honoraria from Novartis, Pierre Fabre and Roche outside the submitted work.

SF reports consulting or advisory board membership: Bayer, Illumina, Roche; honoraria: Amgen, Eli Lilly, PharmaMar, Roche; research funding: AstraZeneca, Pfizer, PharmaMar, Roche; travel or accommodation expenses: Amgen, Eli Lilly, Illumina, PharmaMar, Roche.

JSU is on the advisory board or has received honoraria and travel support from Amgen, Bristol Myers Squibb, GSK, Immunocore, LeoPharma, Merck Sharp and Dohme, Novartis, Pierre Fabre, Roche, Sanofi outside the submitted work

ME has received honoraria and travel expenses from Novartis and Immunocore

SHo received travel support for participation in congresses, (speaker) honoraria and research grants from Almirall, UCB, Janssen, Novartis, LEO Pharma and Lilly outside of the present publication.

SP has received travel support for participation in congresses and/or speaker honoraria from Abbvie, Lilly, MSD, Novartis, Pfizer and Sanofi outside of the present publication.

SPod is on the advisory board or has received honoraria from Galenicum Derma, ISDIN, Cantabria Labs and Mesoestetic.

RLB has received support from Castle Bioscience for the International Melanoma Pathology Study Group Symposium and Workshop.

MG served as consultant to argenx (honoraria paid to institution) and Almirall and received honoraria for participation in advisory boards / travel support from Biotest, GSK, Janssen, Leo Pharma, Lilly, Novartis and UCB - all outside the scope of the submitted work.

MVH received honoraria from MSD, BMS, Roche, Novartis, Sun Pharma, Sanofi, Almirall, Biofrontera, Galderma.

KG declares no competing interests.

**Funding**

This study was funded by the Ministry of Social Affairs, Health and Integration of the Federal State Baden-Württemberg, Germany (grant: AI-Translation-Initiative ("KI-Translations-Initiative"); grant holder: TJB, German Cancer Research Center, Heidelberg, Germany).

# Abstract

Although artificial intelligence (AI) systems have been shown to improve the accuracy of initial melanoma diagnosis, the lack of transparency in how these systems identify melanoma poses severe obstacles to user acceptance. Explainable artificial intelligence (XAI) methods can help to increase transparency, but most XAI methods are unable to produce precisely located domain-specific explanations, making the explanations difficult to interpret. Moreover, the impact of XAI methods on dermatologists has not yet been evaluated.

Extending on two existing classifiers, we developed an XAI system that produces text- and region-based explanations that are easily interpretable by dermatologists alongside its differential diagnoses of melanomas and nevi. To evaluate this system, we conducted a three-part reader study to assess its impact on clinicians' diagnostic accuracy, confidence, and trust in the XAI-support.

We showed that our XAI's explanations were highly aligned with clinicians' explanations and that both the clinicians' trust in the support system and their confidence in their diagnoses were significantly increased when using our XAI compared to using a conventional AI system. The clinicians' diagnostic accuracy was numerically, albeit not significantly, increased. This work demonstrates that clinicians are willing to adopt such an XAI system, motivating their future use in the clinic.

# Main

Melanoma is responsible for most skin cancer-related deaths worldwide. Early detection and excision are critical for ensuring that patients achieve the best prognosis[1]. However, early melanomas are difficult to distinguish from other skin tumours. Recent advances in artificial intelligence (AI)-based diagnostic support systems in dermatology[1,2] have allowed dermatologists to diagnose melanoma and nevi more accurately with AI support when shown digitised images of suspicious lesions.

While this is a promising development, the evaluation of AI support in clinical practice is substantially impeded by the fact that DNNs lack transparency in their decision making[3]: the General Data Protection Regulation (GDPR) requires all algorithm-based decisions to be interpretable by the end users[4]. Additionally, as found in a recent survey by Tonekaboni et al.[5] clinicians want an understanding of the subset of characteristics that determines a DNN's output. As the ultimate responsibility for a diagnosis lies with the clinician, informed clinicians will justifiably be cautious of employing DNN-based systems without being able to comprehend their "reasoning", as DNNs tend to incorporate all correlated features into their decision-making, including spurious correlations[6,7]. Thus, addressing the intransparency of DNNs will allow researchers to comply with the EU Parliament recommendation that future AI algorithm development should involve continual collaborations between AI developers and clinical end users[8,9]

To address the inadequacies of DNN models a variety of explainable artificial intelligence (XAI) methods, i.e. methods aiming to make the reasoning of AI systems more transparent, have been proposed[10–16]. The two primary branches of XAI techniques are (1) inherently interpretable algorithms that are designed to be intrinsically understandable and (2) post hoc algorithms that are designed to retrospectively explain the decisions from a given DNN. However, inherently interpretable algorithms often come with an unfavourable performance-interpretability trade-off[17]. On the other hand, post hoc XAI methods are often rejected as solutions to the problem of transparency in the medical domain due to the risk of introducing confirmation bias when interpreting the explanations[18,19].

Rudin[19] objects to the use of post hoc XAI methods for high-stakes decisions, as they typically require the user to interpret the explanations, that is, to interpret *why* a highlighted image region is relevant or *where exactly* a relevant feature is located - we refer to this issue as the interpretability gap of XAI. The interpretability gap raises a subtle but critical issue of confirmation bias[20] and thus decreases the trust of informed users in an XAI system in which this gap has not been closed. For example, if the XAI diagnoses a melanoma based on the lower left part of the lesion, the underlying human expectation is that there is a genuine diagnostically relevant feature in the lower left of the lesion. If the dermatologist finds a relevant feature in this area upon closer inspection, they will assume that the machine decision was based on this feature, although they cannot be sure of it since they only know *where* the XAI paid attention, but not *why*. The significance of this issue becomes even more apparent if the XAI-highlighted image region does not contain any genuine diagnostically relevant features, leaving the user clueless as to what the XAI decision was based on. Similarly, if the highlighted region contains not only diagnostically relevant features but also features that are spuriously correlated with the diagnosis, such as rulers or surgical skin marks[6,7], the user also cannot be sure whether the diagnosis was made due to genuine features and thus is reliable.

Ghassemi et al.[18] have argued that thorough internal and external validation can be used to address concerns about the reliability of DNN models while eschewing the issues introduced by XAI. Although we agree that thorough validation is important, a solution to the transparency problem that closes the interpretability gap and therefore allows users to build (justifiable) trust in a reliable support system is crucial to improve clinicians' diagnostic performance and thus further improve the patient outcome.

We therefore aimed to develop a multimodal XAI that provides explanations that (1) close the interpretation gap by localising well-established dermoscopic features[21–23] such that (2) it can be easily interpreted by dermatologists in the initial diagnosis of melanoma to (3) assess how clinicians interact with the XAI. Building on prior work[24,25], we developed and evaluated our multimodal XAI (Fig. 1a) on a novel dataset containing expert annotations. We evaluated our XAI's influence on clinicians in a three-phase reader study with 116 international participants (Fig. 1b). We found that our XAI achieves good diagnostic performance on the distinction between melanoma and nevus. Its explanations for the task are well-aligned with those used by human experts. While our XAI did not improve the diagnostic accuracy of the clinicians it increased their trust in their own diagnosis and the trust in the support system compared to non explainable AI support. We showed that the increase in trust is correlated with the overlap between human and machine explanation. Finally, we published our dermatologist-annotated dataset to encourage further research in this area. Thus, we took an important step towards the employment of XAI in the clinic to improve patient outcomes by developing a trustworthy well-performing XAI support system that complies with the GDPR, EU recommendations and clinicians' expectations of AI support systems[4,9,26].

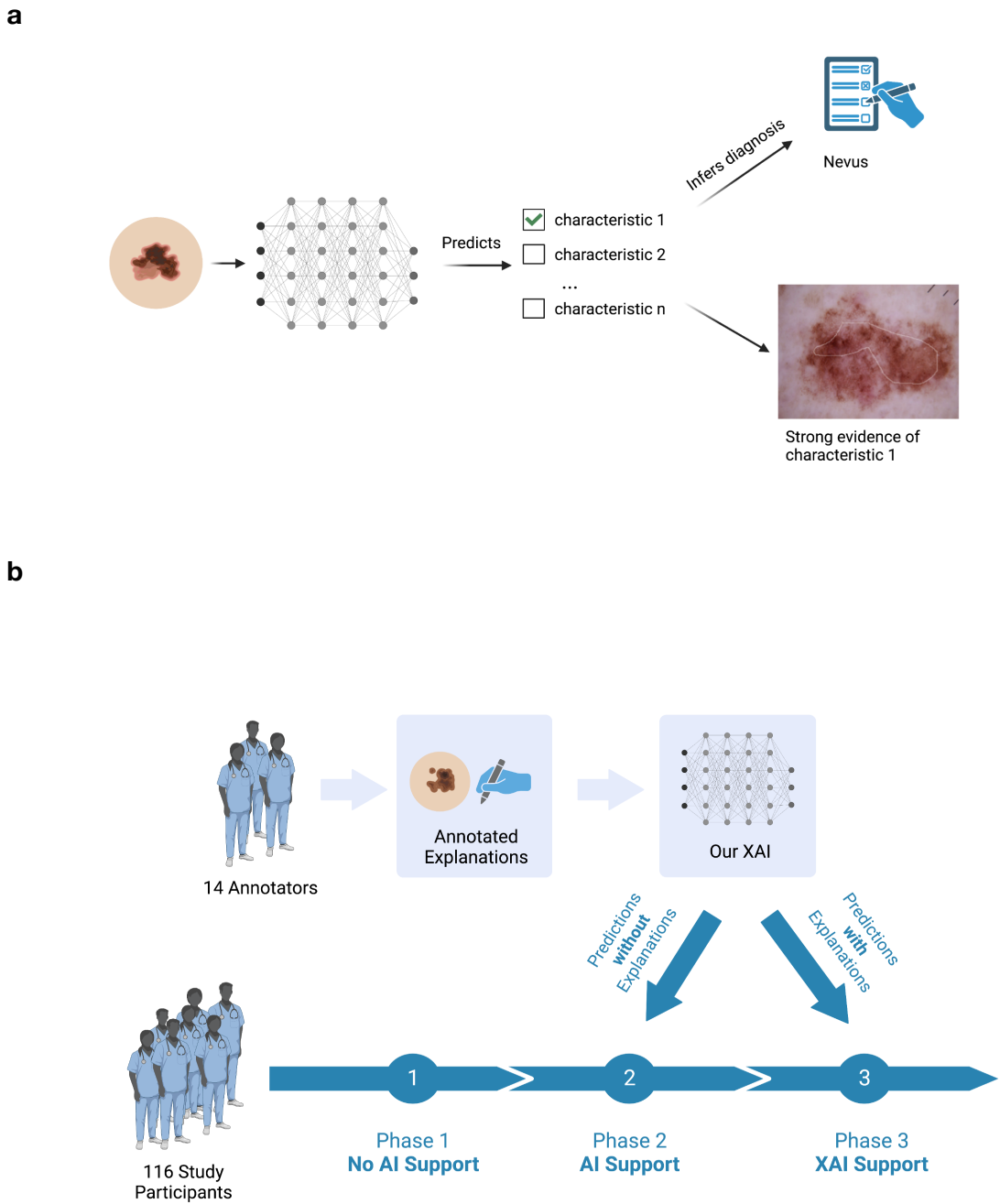

**Fig. 1: Overview of the XAI and reader study.**

**a**, Schematic overview of our multimodal XAI. The AI system makes a prediction for each characteristic and then infers a melanoma diagnosis if it detects at least two melanoma characteristics. The diagnosis and corresponding explanations are then displayed to the clinician. **b**, Schematic overview of our work. We first collected ground-truth annotations and corresponding ontology-based explanations for 3611 dermoscopic images from 14 international board-certified dermatologists and trained an explanatory AI on this dataset (top row). We then employed this classifier in a three-phase study (bottom row) involving 116 clinicians tasked with diagnosing dermoscopic images of melanomas and nevi. In phase 1 of the study, the clinicians received no AI assistance. In phase 2, they received the XAI's predicted diagnoses but not its explanations. In phase 3, they received the predicted diagnoses along with the explanations. Figures created with BioRender.com.

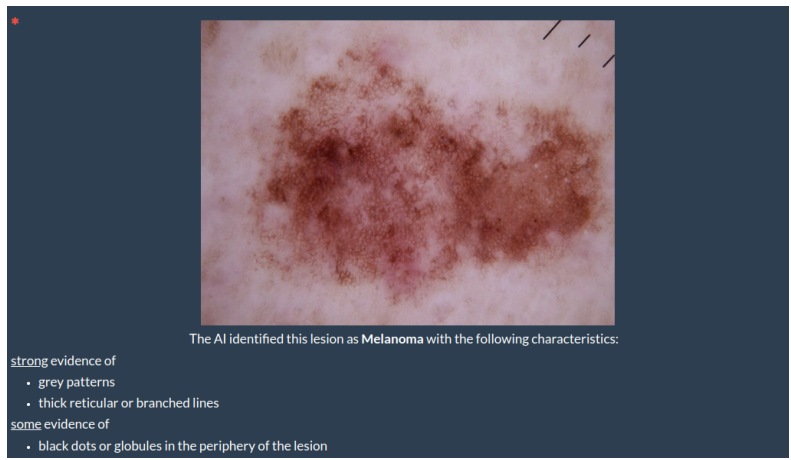 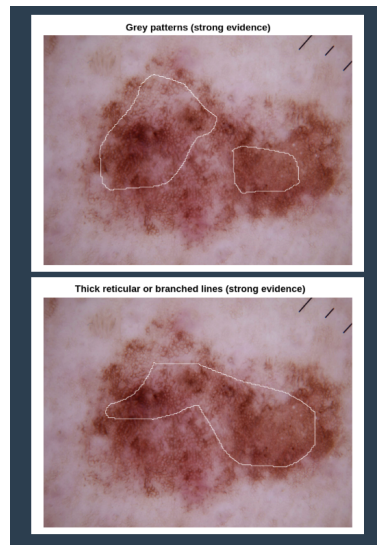

**Fig. 2: Example multimodal XAI explanation.**

**a, b**: An example multimodal explanation from our XAI used in phase 3, showing a textual explanation (**a**) and the corresponding localised visual explanations (**b**). The XAI identified this lesion as a melanoma with the characteristics stated in the textual explanation. The white polygons represent the most important regions where the XAI detected the corresponding characteristics.

# Results

## Our XAI achieves good diagnostic accuracy

We first evaluated the diagnostic performance of our XAI compared to a baseline state-of-the-art classifier (Supplementary Information E) to ensure that the XAI does not compromise diagnostic accuracy. The balanced accuracy of our XAI on the test set was 81% (95% CI: [75.6, 86.3%]) while the baseline classifier achieved 80% (95% CI: [74.4%, 85.4%]) Additionally, we assessed the extent to which our model learned to pay attention to the lesion, as the surrounding skin does not contain meaningful information for the differential diagnosis between melanoma and nevus. To this end, we computed the ratio of the mean Grad-CAM[12] attribution value within the lesion to that outside of the lesion. Our XAI focused on regions inside the lesion significantly more than the baseline classifier (*P*<0.0001, two-sided Wilcoxon signed-rank test, n=196 images), with ratios of 35.9 (95% CI: [30.7, 42]) versus 4.1, respectively (95% CI: [3.4, 4.7]) (Fig. 3a). Robustness in the presence of artefacts is illustrated in Extended Data Fig. 1. Thus, our XAI provides additional interpretability while maintaining the same level of diagnostic accuracy.

## Our XAI is strongly aligned with clinicians' explanations

In phase 1 of the study, participating clinicians were asked to select explanations from an explanatory ontology and localise them on the lesion during diagnosis. We used these explanations to determine the extent to which the XAI system and clinicians detected similar explanations on the same lesions. We evaluated our XAI's alignment with the clinicians' ontological explanations and annotated regions of interest (ROI) by assessing their overlap.

To assess the overlap in ontological explanations, we calculated the Sørensen-Dice similarity coefficients (DSC)[27] between the ontological characteristics selected by the clinicians in phase 1 and those predicted by our XAI. The Sørensen-Dice similarity ranges from 0 to 1, where 0 indicates no overlap and 1 indicates full overlap. The mean explanation overlap was 0.46 (95% CI: [0.44, 0.48], 366 images) when both the clinicians and the XAI predicted melanoma and 0.23 (95% CI [0.20, 0.26], 505 images) when they both predicted a nevus. Considering both diagnoses, the mean explanation overlap was 0.27 (95% CI: [0.25, 0.29], 1089 images) (Fig. 3b). For comparison, we assessed between-clinician overlap to determine the level of agreement among clinicians for the same images. We computed the DSC for each pair of clinician-selected ontological characteristics per image, as each image was diagnosed by multiple clinicians. The mean between-clinician overlap was 0.28 (95% CI: [0.27, 0.29], 5165 image pairs), which is comparable to the overlap between the AI and clinicians.

Next, we investigated the overlap between human and machine ROIs. We defined human ROIs as the image regions that the clinicians annotated to explain their diagnoses in phase 1. For the machine ROIs, we computed the gradient-weighted class activation maps (Grad-CAMs[12]) for the same images, i.e., the image regions that had the biggest influence on the machine's diagnosis. We defined the ROI overlap for an image as the DSC between both sets of ROIs. The mean ROI overlap attained by our XAI was 0.48 (95% CI: [0.46, 0.5]) (Fig. 3b, c). For comparison, we performed the same calculations using the baseline classifier and compared them to the overlap of our XAI. The baseline classifier achieved a mean overlap of 0.39 (95% CI: [0.38, 0.41]). Thus, our XAI yielded significantly higher ROI overlap than the baseline ($P<0.0001$, two-sided paired t test, n=1120 images)) (Fig. 3a; Extended Data Fig. 2a, b).

Thus, the XAI system shows strong alignment with clinicians on both the ontological explanations and the ROI modalities.

## AI support improves diagnostic accuracy, but XAI support does not further increase diagnostic accuracy over AI support alone

We assessed our XAI's influence on the clinician's diagnostic accuracy compared to both plain AI support and no AI support. To investigate the relationship between the clinicians' experience levels and benefit with XAI over AI support, we correlated their change in accuracy with their reported experience in dermoscopy.

To differentiate the effect of XAI from the effect of receiving AI support, we first investigated the influence of AI support on the clinicians' diagnostic accuracy against not receiving any AI support (phase 1). Out of 109 participants, we observed a performance improvement with AI support for 77 participants, a decrease for 31, and no change for 1. The clinicians' mean balanced accuracy was 66.2% (95% CI: [63.8%, 68.7%]) in phase 1 and 72.3% (95% CI: [70.2%, 74.3%]) in phase 2 (Fig. 4a). Pairwise comparison revealed a statistically significant improvement ($P=P<0.0001$, two-sided paired t test, n=109 participants) by AI support alone.

Next, we investigated whether XAI support had an effect on the clinicians' diagnostic accuracy beyond the benefits of receiving AI support. We compared the clinicians' balanced accuracies between phases 2 and 3 and observed a performance improvement with XAI support for 52 participants, a decrease for 34, and no change for 30. The participants' mean balanced accuracy was 73.2% (95% CI: [71%, 75.3%]) with XAI (Fig. 4a), a slight increase from phase 2. However, a pairwise test revealed that the difference was not significant ($P=0.34$, two-sided paired t test, n=116 participants). The full details are provided in Extended Data Tab. 1.

We observed a slight correlation (correlation coefficient 0.2, 95% CI: [0.02, 0.37], $P=0.03$, Spearman's rank correlation, n=116 participants) between the clinicians' reported experience in dermoscopy and their increase in accuracy with XAI over AI. The clinicians that reported involvement in regular scientific discussions about dermoscopy gained more from XAI support than from AI alone. The group that showed the highest accuracy with AI support alone and also showed the largest decline in accuracy with the use of the XAI reported that they rarely performed dermoscopy (Fig. 4d).

Further analysis revealed that the clinicians were slightly more likely to agree with the machine decisions made in phase 3 than with those made in phase 2. We determined the clinicians' agreement with the AI/XAI by calculating the proportion of images where the human diagnoses matched the machine's predicted diagnoses. The mean agreement was 79.5% (95% CI: [77.1%, 81.2%]) with the XAI in phase 3 versus 77.1% (95% CI: [75%, 79.2%]) agreement with the AI in phase 2. Pairwise testing showed that the mean percentage point increase of 2.4% (95% CI: [0.65% 4.2%], $P=0.009$, two-sided paired t test, n=116 participants) was significant. We also computed the clinicians' agreement specifically for cases in which the support system was wrong. We observed that the clinicians' agreement on erroneous predictions also increased from 63% (95% CI: [57%, 69.1%]) in phase 2 to 67.9% (95% CI: [61.9%, 73.7%]) in phase 3. However, the pairwise mean increase of 4.8 (95% CI: [-1.2, 10.9]) percentage points was not statistically significant ($P=0.126$, two-sided paired t test, n=116 participants).

Thus, AI support increased clinician diagnostic accuracy but XAI support did not significantly improve it further, despite the higher agreement. Clinicians with the most experience in dermoscopy benefited most from XAI support, whereas clinicians with less experience benefited most from plain AI support.

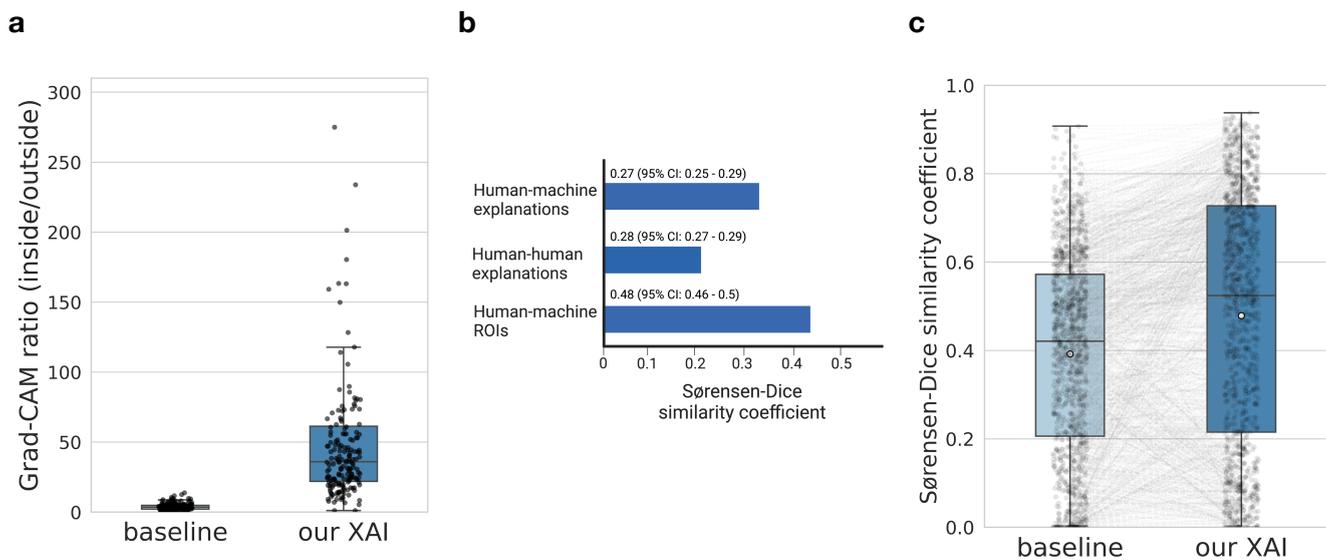

**Fig. 3: Overview of our XAI's characteristics**

**a**, Ratio of mean Grad-CAM pixel activation value inside the lesion to that outside the lesion. Higher values are better, as they indicate greater attention on regions within the lesion than on regions outside the lesion. Four data points for the baseline and 19 data points for the XAI have values above 300 and have been omitted to more clearly visualise the data. **b**, Overlap between human and machine reasoning in both regions of interest and explanations. **c**, Comparison between our XAI's Region of interest (ROI) overlap between humans and AI and that of the baseline. The central line on each box denotes the median value. The upper and lower box limits denote the 1st and 3rd quartiles, respectively, and the whiskers extend from the box to 1.5 times the interquartile range. Some figures created with BioRender.com.

# XAI significantly increases clinicians' confidence in their own diagnoses

To compare the influence of AI and XAI support on the confidence clinicians had in their own diagnosis, we compared the participants' confidence scores for each image between phases 2 and 3 for each image. We observed a mean increase of 12.25% (95% CI: [9.06%, 15.74%]) in confidence with XAI support relative to only AI support ($P<0.0001$, two-sided paired t test, n=1714 images). The mean confidence per participant in each phase is illustrated in Fig. 4b and provided in detail in Extended Data Tab. 2. The absolute confidence values for the images are reported in Extended Data Fig. 3a.

Next, we analysed the influence of displaying the confidence of the XAI on the clinicians' own diagnostic confidence. In phase 3, we observed a slight difference in the clinicians' confidence when the classifier was confident versus when it was uncertain. The mean human confidence score for high-confidence AI predictions was 7.82 (95% CI: [7.73, 7.91]), and that for uncertain AI predictions was 7.69 (95% CI: [7.56, 7.81]) ($P=0.039$, two-sided Mann–Whitney U test, $n_{high}$=1218 images, $n_{low}$=496 images). In phase 2, when the participants received no information about the classifier's certainty, this disparity in the clinicians' confidence was statistically insignificant, with mean confidence scores of 7.54 (95% CI: [7.44, 7.64]) for high-confidence AI predictions and 7.47 (95% CI: [7.31, 7.61]) for low-confidence AI predictions ($P=0.319$, two-sided Mann–Whitney U test, $n_{high}$=1218 images, $n_{low}$=496 images).

In conclusion, the clinicians' confidence in their diagnoses significantly increased with XAI support relative to plain AI support. The XAI's communicated confidence was also correlated with the clinicians' own confidence in their decisions.

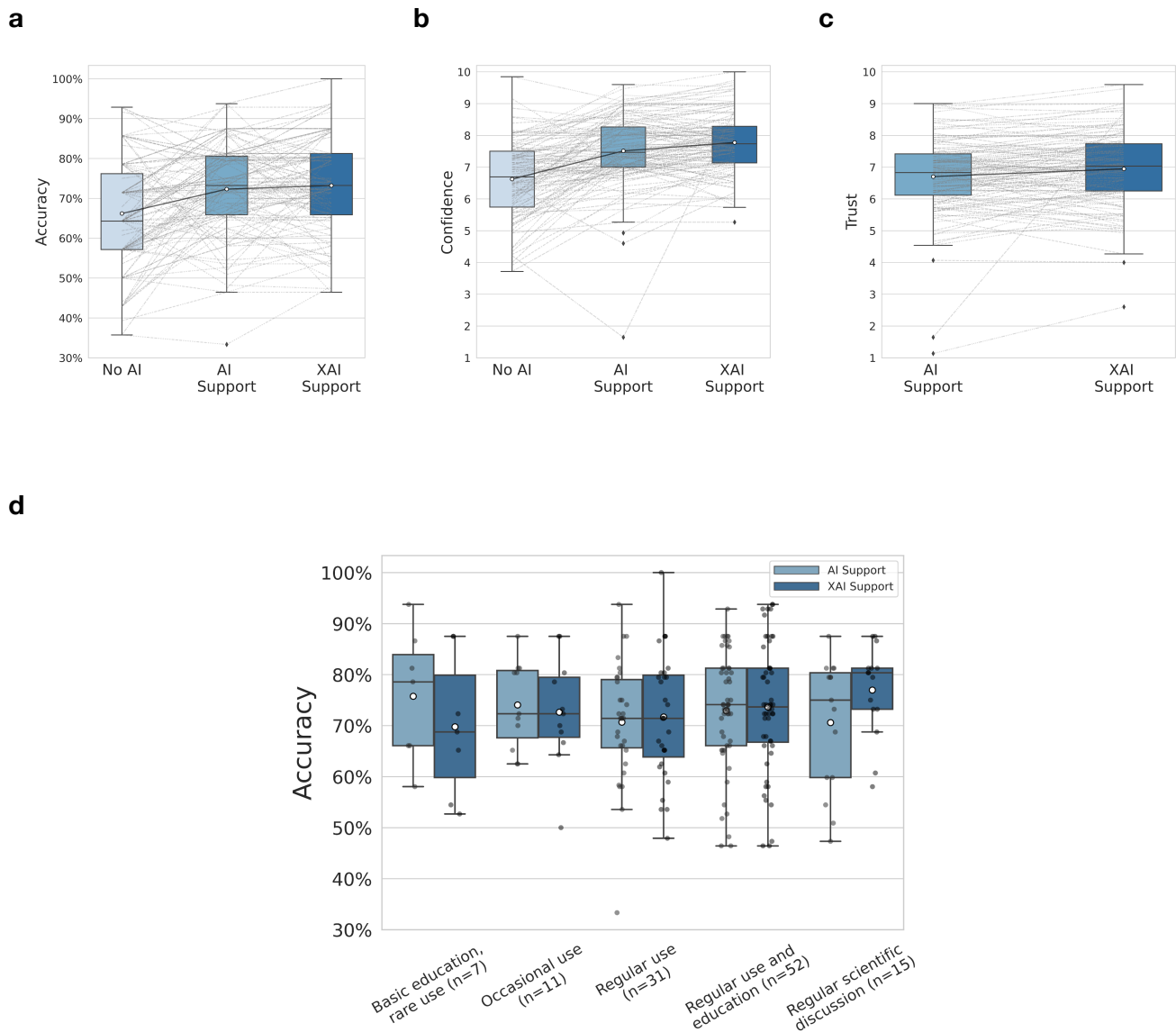

**Fig. 4: Impact of our XAI on clinicians's diagnostic accuracy, confidence, and trust.**

**a-c**: Distributions of clinicians' mean balanced accuracy (**a**), confidence (**b**), and trust (**c**) in each phase of our study. The grey lines between the phases connect the same participant between phases, and the black lines connecting the boxes indicate the means across all participants. The central line on each box denotes the median value. The upper and lower box limits denote the 1st and 3rd quartiles, respectively, and the whiskers extend from the box to 1.5 times the interquartile range. **d**, Balanced diagnostic accuracy with AI and XAI support grouped by different levels of experience with dermoscopy. The clinicians that reported rarely using dermoscopy saw a decrease in balanced accuracy, while clinicians who were involved in regular scientific discussion (papers, conferences, etc.) saw the highest accuracy increase with XAI.

## XAI significantly increases clinicians' trust in machine decisions

To assess the impact of the XAI's explanations on the clinicians' trust in the AI's decisions, we compared the trust scores between phase 2 and phase 3 for each image. The mean increase in trust with XAI support in phase 3 was 17.52% (95% CI: [13.74%, 21.6%]), and a pairwise comparison revealed a statistically significant increase relative to only AI support in phase 2 ($P<0.0001$, two-sided paired t test, n=1714 images). We also observed that trust scores in both phases were significantly dependent on whether or not the clinicians agreed with the AI diagnoses [7.55 (95% CI: [7.48, 7.62]) when they agreed vs. 4.8 (95% CI: [4.64, 4.96]) when they disagreed, $P<0.0001$, two-sided unpaired t test, $n_{agreed}$=2684 images, $n_{disagreed}$=744 images]. The mean trust per participant in each phase is illustrated in Fig. 4c and provided in detail in Extended Data Tab. 2 and the absolute trust values are given in Extended Data Fig. 3b.

Next, we analysed the influence of displaying the confidence of the XAI on the clinicians' trust in the machine predictions. In phase 3, the classifier's transparency about its low certainty did not affect the participants' trust scores. The mean trust score was 6.95 (95% CI: [6.77, 7.14]) for high-confidence AI predictions and 6.96 (95% CI: [6.77, 7.14]) for low-confidence AI predictions ($P=0.6$, two-sided Mann–Whitney U test, $n_{high}$=1218 images, $n_{low}$=496 images). However, we found the same effect in phase 2, where the participants received no AI confidence information: the mean trust score was 6.74 (95% CI: [6.6, 6.87]) on high-confidence AI predictions and 6.67 (95% CI: [6.47, 6.86]) on low-confidence AI predictions ($P=0.42$, two-sided Mann–Whitney U test, $n_{high}$=1218 images, $n_{low}$=496 images).

Hence, clinicians' trust in the machine decisions significantly increased with XAI support relative to plain AI support. Also, the XAI's communicated confidence did not affect the clinicians' trust in the XAI's predictions.

## Clinicians' trust in AI decisions is slightly correlated with overlap in ontological explanations when the clinician and AI diagnoses match

We hypothesised that clinicians' trust in an AI system is correlated with the amount of overlap between their ontological explanations and the machine's ontological explanations. To determine this, we investigated the correlation between clinicians' trust in AI and the overlap in ontological explanations between the clinicians and the AI. We again defined overlap in ontological explanations as the Sørensen-Dice similarity between the clinician-selected ontological characteristics determined in phase 1 and the XAI-predicted characteristics. To isolate the influence of overlapping explanations on the clinicians' trust in the AI diagnoses, we calculated the overlap and trust correlations by considering only the images where the clinicians' diagnoses matched the AI's diagnoses.

When the clinicians and AI agreed, we observed a slight correlation between trust and overlap in reasoning (correlation coefficient 0.087, 95% CI: [0.02, 0.15], $P=0.01$, Spearman's rank correlation, n=871 images) (Fig. 5a). As a sanity check, we assessed the correlation again using the phase 2 trust scores instead. Since no explanations were shown in phase 2, we expected that there would be no discernible correlation between explanation overlap and trust scores. This was indeed the case, as the correlation coefficient was -0.05 (95% CI: [-0.1, 0.01], P=0.097, Spearman's rank correlation, n=866 images).

Upon further investigation, we noticed a difference in the distribution of overlap between the two diagnoses. When both the clinicians and the AI predicted melanoma, the correlation coefficient between trust and the overlap in reasoning was 0.23 (95% CI: [0.19, 0.34], *P*<0.0001, Spearman's rank correlation, n=567 images), and when both predicted nevus, the correlation coefficient was -0.1 (95% CI: [-0.19, -0.02], *P*=0.01, Spearman's rank correlation, n=505 images) (Fig. 5b, c).

Thus, we found that clinicians place higher trust in machine decisions when their reasoning overlaps. However, this effect was only observable when they diagnosed melanoma compared to when they diagnosed nevus.

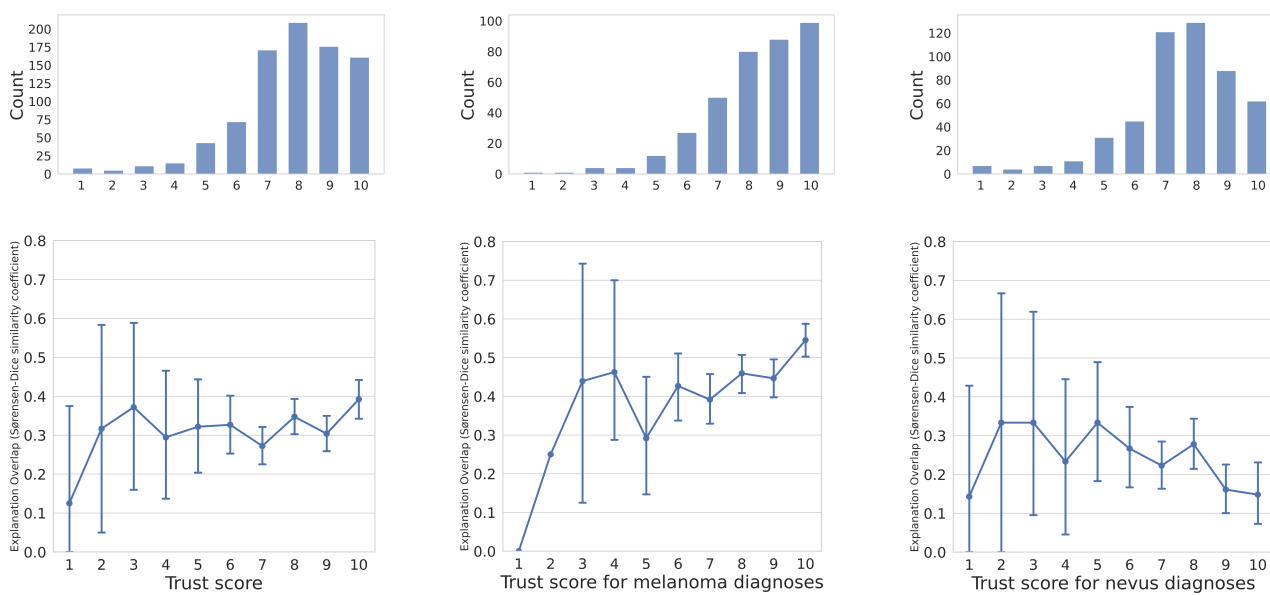

**Fig. 5: Relationship between clinicians' trust in AI and overlap in ontological explanation when the diagnoses match.**

**a-c**: Correlation between overlap in reasoning and trust in AI for cases where the clinicians' diagnoses matched those of the AI. On the bottom row, the filled circles denote the means, and the error bars represent the 95% bootstrap CI. The left column depicts the relationship between overlap (measured by Sørensen-Dice similarity coefficient) in reasoning and trust in AI for both classes (**a**), the middle column depicts cases where both the clinicians and the AI diagnosed melanoma (**b**), and the right column represents cases where they both diagnosed nevus (**c**). The histograms above the plots depict the distributions of the trust scores.

# Discussion

Our work intends to close the interpretability gap in AI-based decision support systems by developing an XAI that can produce domain-specific interpretable explanations to aid in melanoma diagnosis. Additionally, we aimed to evaluate the XAI system's effect on clinicians' diagnostic accuracy, confidence, and trust in the system, and to assess the factors that contribute to trust. To this end, we designed a multi-modal XAI system with clinical explanations and conducted the first large-scale reader study on such an XAI system in dermatology.

In our work, we showed that the diagnostic performance of our XAI was on par with the baseline, while being interpretable by learning human-relevant features. Additionally, we demonstrated that our XAI was minimally affected by spurious correlations in two ways. First, we quantitatively showed that our XAI was well aligned with clinician-relevant ROIs. Aligning classifier ROIs with clinician ROIs provides a first reassurance that the AI is not making predictions based on spurious correlations. Second, common artefacts that can alter the output score of DNNs used for dermatology[7,28] are almost always located in the area surrounding the lesion and rarely within it. Therefore, we assessed the average pixel attributions within the lesions versus those around them and found that this ratio was significantly greater for our XAI than for the baseline classifier, i.e. we found that the baseline classifier learned false associations significantly more often than our XAI. Hence, we refute the widely accepted notion of performance-interpretability trade-off, and consequently, future AI development can emphasise learning human-relevant features.

We found that AI support improved clinicians' diagnostic accuracy over no AI support, but XAI did not further increase clinicians' diagnostic accuracy compared to AI support alone. The increase in accuracy with AI support is consistent with other studies[1,2]. Therefore, AI assistance can considerably improve diagnostic accuracy, but the effects of XAI still need to be fully explored. Since we observed that experienced clinicians benefit most from XAI and inexperienced ones from plain AI, future research should investigate how experienced versus inexperienced clinicians interact with an XAI as well as the effects of erroneous AI predictions.

We observed that clinicians' confidence significantly increased with AI support and that XAI support enhanced this effect even further. A previous study found that clinicians' confidence did not increase with AI support[1], while another study found an increase in 11.8% of the cases[29]. The non-increase in the first study could be attributed to the fact that the participants were shown uncalibrated confidence scores of the AI, which is known to be overconfident even when making incorrect predictions[30]. Seeing the AI exhibit strong confidence in the majority of cases where it may have been incorrect or ambiguous likely prevented an increase in their diagnostic confidence. In contrast, we presented calibrated confidence scores to the participants, which may have had a more favourable effect on their diagnostic confidence. As a result, XAI can increase the adoption of AI assistants since it enhances clinicians' confidence in their decisions.

We showed that clinicians place more trust in a support system's diagnoses when the system explains the reasoning behind its decisions in a way that is interpretable. This aligns with the recent survey by Tonekaboni et al.[26] where clinicians emphasised the need for interpretability. Additionally, we showed that the clinicians' trust in the XAI was correlated with the overlap between the XAI's and clinicians' reasoning. However, while this correlation was significant for images classified as melanoma, we did not observe the same effect for those classified as nevus. When predicting a nevus, "melanoma simulator" was the XAI's most common explanation, as one ground-truth annotator selected this explanation for the majority of nevus images. Many of the participants did not concur with this explanation, which may explain the lack of overlap-trust correlation for the nevus images. A possible explanation can be found in the work of Grgić-Hlača et al.[31] who found that seeing a machine decision assistant make errors similar to human errors influences human advice-taking behaviour and that humans perceive similar machine assistance as more accurate, useful, and predictable. When explaining nevus predictions, the XAI made mistakes that humans wouldn't make, i.e. the "melanoma simulator" explanation, which may account for the lack of correlation in the nevi predictions. We believe additional research on this topic will be beneficial to fully understand the factors that influence clinicians' trust in AI. Nonetheless, XAI can increase clinicians' trust in AI assistance, and as they are more likely to follow the advice of a system they trust, this can result in increased adoption.

A limitation of our work is that similar to prior works[32], our system was tested under artificial conditions. Furthermore, in contrast to common post-hoc XAI methods, which are intended to make the AI's inner workings transparent without changing the model itself, we intentionally guide our system to provide human-like explanations. This decision comes with the cost of sacrificing the AI's potential ability to recognize patterns that humans do not observe. Nonetheless, we believe that in safety-critical settings, such as melanoma detection, providing explanations that users are familiar with and that they can consequently verify yields a comparably larger benefit. Additionally, our work does not deal with the domain shift problem, i.e. a situation in which the data encountered after deployment in the clinic is significantly different from the data the AI was trained on. The performance of our classifier must be extensively verified in domain shift scenarios, with the addition of unsupervised domain-adaptation techniques[33], such as Adversarial Residual Transfer Networks[34]. In the future, we plan to expand on our ground-truth dataset and include images from multiple clinics, which may improve the generalizability of our XAI to multiple clinics.

Our work advances the field of human-computer collaboration in medicine by developing a multimodal XAI that generates human-like explanations to support its decisions and evaluating its influence on clinicians. The EU Parliament recommends that future AI algorithm development should involve continual collaborations between AI developers and clinical end users[8,9]. Our XAI design was guided by the end user perspective and can iteratively be improved with the results of this study, further clinician feedback, and future work on XAI-human collaboration. The findings of our reader study illustrate the ability of XAI to enhance clinicians' diagnostic confidence and furthermore suggest that it has the potential to increase the likelihood of considering machine suggestions. The European General Data Protection Regulation (GDPR) requires that all algorithm-based decisions be interpretable by the end users[4]. As our work addresses the current discrepancy between legal and ethical requirements for transparent clinical XAI systems and state-of-the-art classifiers, it constitutes an important first step towards closing the interpretability gap.

# Methods

Prior to data collection, we registered our hypotheses and analysis plan on the Open Science Framework[1] website, which may be accessed under https://osf.io/g3keh. We followed the STARD guidelines[35], which we report in detail in Supplementary Information A.

## Explanatory ontology

To allow for speedy annotation by both our annotating dermatologists as well as the study participants in phase 1 of the reader study and to facilitate a streamlined evaluation of the explanations, we created an ontology containing typical features of melanomas and nevi based on pattern analysis. We combined well-established features from several sources[21–23] and included feedback from the dermatologists participating in the pilot study (see Supplementary Information B for details) as well as from the ground-truth annotators. The ontology was first compiled in German and approved by a panel of board-certified dermatologists prior to the study. The German-speaking participants of the three-phase reader study received the original German version of the ontology. We translated the ontology to English for the international study participants. The translation was approved by two board-certified dermatologists (RB, MLV). All features are listed in Tab. 2. More details on the ontology can be found in Supplementary Information C.

| **Melanoma criteria** | **Nevus criteria** |
|---|---|
| <ul><li>Thick reticular or branched lines</li><li>Eccentrically located structureless area (any colour except skin colour, white and grey)</li><li>Grey patterns</li><li>Polymorphous vessels</li><li>Pseudopods or radial lines at the lesion margin that do not occupy the entire lesional circumference</li><li>Black dots or globules in the periphery of the lesion</li><li>White lines or white structureless area</li><li>Parallel lines on ridges (acral lesions only)</li><li>Pigmentation extends beyond the area of the scar (only after excision)</li><li>Pigmentation invades the openings of hair follicles (facial lesions)</li></ul> | <ul><li>Only one pattern <u>and</u> only one colour</li><li>Symmetrical combination of patterns and/or colours</li><li>Monomorphic vascular pattern</li><li>Pseudopods or radial lines at the lesional margin involving the entire lesional circumference</li><li>Parallel lines in the furrows (acral lesions only)</li><li>Pigmentation does not extend beyond the area of the scar (only after excision)</li><li>Asymmetric combination of multiple patterns and/or colours in the absence of other melanoma criteria</li><li>Melanoma simulator</li></ul> |

**Tab. 1: Melanoma and nevus criteria used in our ontology.** With the exception of the feature "melanoma simulator", the nevus criteria always imply the absence of distinct melanoma features.

---

[1] https://osf.io/

## Images and annotation procedure

We used the publicly available dataset HAM10000[36] for our study, which contains 10015 dermoscopic images of several skin diseases at different localizations on the body. The dataset contains images from both sexes and patient age ranges from 0 to 85 reported in 5-year intervals. During the construction of the dataset, several images per lesion were often taken, and occasionally, more than one lesion per patient was included. Thus, the number of images is greater than the number of unique lesions, and the number of unique lesions is greater than the number of patients. The diagnoses were confirmed by excision and subsequent pathological evaluation, by panel decision or by follow-up.

In this study, we used all the biopsy-verified melanoma and nevus images in the HAM10000 dataset, i.e., a set of n=3611 images of 1981 unique lesions. We refer to this set of images as the "base set" in the remainder of this section. To acquire the necessary annotations for training the classifier, we asked 14 international board-certified dermatologists to annotate these 3611 images of biopsy-verified melanomas and nevi from the dataset. To prevent any data loss as a result of misdiagnoses, we provided the ground truth diagnoses to the annotators. With knowledge of the diagnosis of the lesion, the annotating dermatologists were tasked with explaining the given diagnosis by selecting the relevant features from the explanatory ontology and by annotating the image regions (ROIs) corresponding to the selected features. One dermatologist (SHo) annotated all 3,611 images, while each of the other 13 annotators annotated between 200 and 300 unique lesions such that our dataset contains annotations by at least two dermatologists per unique lesion. We set the explanatory labels for an image as the union of the annotator's explanatory labels, and we merged the ROIs according to the information loss of the merged ROI relative to the original ROIs (full details can be found in Supplementary Information D).

We split the base set into a training set, a validation set and a test set. The test set contained 200 unique lesions with 100 unique melanomas and nevi each. For this, we randomly chose 100 unique melanomas and nevi (with complete information on patient age and sex, as well as on the localization of the lesion) from the base set. For the test set, we kept only one image per lesion; in cases where several images were present for a single lesion, we chose the last image as identified by the image ID. After assigning images to the test set, we proceeded by removing all images that contained the selected lesions as well as other lesions from the same patients from the base set. We then performed a random 82:18 split on the unique lesions in the remainder of the base set to form the training set and the validation set, respectively. In doing so, we ensured that all images of lesions that were photographed multiple times as well as lesions from the same patient were contained in only one of the sets to avoid leaking information from the training to the validation set. As a result, our training set contained 2646 images of 1460 lesions, and the validation set contained 599 images of 321 lesions. Around 22% of the lesions in each set were melanomas, while 78% were nevi.

# XAI Development

## Classifier design

We used Python version 3.9 and the Pytorch library version 1.12.1 to train our XAI. We developed an AI classifier that is able to explain itself to clinicians by making use of the well-established visual characteristics[21–23] of melanoma and nevi from our explanatory ontology. Our classifier learns to predict these characteristics from digitised dermoscopic images and infers the diagnosis of melanoma or nevus from its predictions.

After acquiring the ground-truth annotations, we trained the classifier to predict the lesion characteristics and the corresponding diagnosis. We follow the attention inference architecture introduced by Li et al[24]. and extended by Jalaboi et al[25]. Our classifier has two components: a classification component $Comp_C$ and a guided attention component $Comp_A$ to help localise the relevant features. In $Comp_C$, instead of predicting the diagnosis directly, the classifier predicts the characteristics from our ontology. We infer the diagnosis as melanoma if at least two melanoma characteristics are detected; empirically, we found that this approach leads to the best trade-off between sensitivity and specificity, and clinically, this approach is similar to the use of the 7-point checklist[22,] which also requires at least two melanoma criteria for a diagnosis of melanoma if used with the commonly used threshold of three points.

To guide the classifier to learn features used by dermatologists and create more meaningful explanations, we employ $Comp_A$. For training, we define the loss $L_A$ in addition to the regular cross-entropy loss between the target and the prediction as:

$$L_A = \frac{1}{N} \sum_{i=1}^{N} (1 - \frac{1}{C} \sum_{c=1}^{C} \frac{2 A_{i,c} H_{i,c}}{A_{i,c} + H_{i,c}})$$

where $N$ is the number of samples, $C$ is the number of classes, $A_C$ is the Grad-CAM[12] attention map generated from the last convolutional layer of $Comp_A$ and $H_C$ is the ground-truth ROI annotated by the dermatologists. For images where the ground-truth label of a characteristic $c$ was 0, i.e., the characteristic was not present in the lesion, we set $H_C$ to be a zero-valued matrix of the same size as $A_C$. This additional loss term was added to the regular cross entropy loss to yield the following combined loss:

$$L = \lambda_C L_C + \lambda_A L_A$$

where $L_C$ is the cross entropy loss for the characteristics and $L_A$ is the Dice loss between the Grad-CAM heatmaps of the model's predictions and the ROIs annotated by the dermatologists. $\lambda_C$ and $\lambda_A$ are hyperparameters for assigning weights to the individual components. For all our experiments, we set $\lambda_C$ to 1 and $\lambda_A$ to 10.

We opted to use a ResNet50 pretrained on the ImageNet dataset[37] as the base model since it has been shown to perform well in skin lesion classification tasks[38]. We used random sampling to balance the class distribution during training. We also used several image augmentations to improve generalizability, in line with the International Skin Imaging Challenge (ISIC) melanoma classification challenge winners, who achieved state-of-the-art performance on a skin lesion classification task[39]. Complete details on the model hyperparameters can be found in Supplementary Information E.

Additionally, we chose to display the confidence of the classifier for each prediction. Conventionally, the raw output of the softmax or sigmoid layer is used as a measure of confidence; however, this value is an unreliable measure of confidence and should be calibrated[30]. To obtain well-calibrated probabilities, we performed temperature scaling on each output class, which is a simple but effective method for calibrating neural network outputs to more accurately reflect model confidence[30].

## Classifier performance testing

We evaluated the performance of the classifier on the held-out test set in terms of balanced accuracy. The sensitivity and specificity of the classifier were determined on the validation set. For the calculation of the ratio of mean Grad-CAM attributions within the lesion to those surrounding the lesion, we used the formula: $Ratio = \frac{\mu(Grad-CAM\ attributions\ within\ lesion)}{\mu(Grad-CAM\ attributions\ surrounding\ lesion)}$. We determined the regions inside and outside of the lesions using the segmentation maps available for the HAM10000 dataset[2].

For the calculation of overlap between the XAI-predicted explanations and the clinician-selected explanations, we used the Sørensen-Dice similarity coefficients (DSC), calculated with the numbers of true positives (*TP*), false positives (*FP*), and false negatives (*FN*) as follows: $DSC = \frac{2TP}{2TP+FP+FN}$. We also used the SC for ROI comparisons, we used the DSC again, this time calculated as $DSC = \frac{2\sum(a \cdot b)+\epsilon}{\sum a + \sum b + \epsilon}$, where $a$ and $b$ are the soft image masks and $\epsilon$ is a smoothing term.

## Design of the explanations

According to the EU Parliament's recommendations regarding AI in healthcare, future AI algorithm development should be based on co-creation, i.e., continual collaborations between AI developers and clinical end users[8,9]. Consequently, we designed our explanation scheme with the consultation of two board-certified dermatologists (SHo, CNG). The explanation scheme included both visual and text-based components as well as assessments of classification confidence.

The majority of AI explanation approaches are visual, using saliency maps to emphasise areas in an image that are most important in making predictions. This is most commonly achieved by superimposing a rainbow-coloured heatmap onto the image, but other visualisations are also possible. These methods were initially created with AI developers' debugging needs in mind. According to the consulting dermatologists, such heatmaps obscured their view of the lesion such that they needed to switch back and forth between the explanation image and the original image without the heatmap. This was deemed tedious and unsuitable for the diagnostic process. The dermatologists expressed the need for a clear view of the lesion in the explanation image, allowing them to quickly determine whether the predicted features are present in the salient regions. Therefore, we decided to indicate the most relevant region(s) for the prediction of each feature by displaying a polygon-shaped ROI over the top 20$^{th}$ percentile attribution values, as shown in Fig. 2b, according to consultation with the consulting dermatologists. Only showing a polygon has the drawback that the person seeing the explanation has to interpret the region within the polygon as the more important region. We experimented with a slight darkening of the unimportant regions outside of the polygon to solve this issue. However, we rejected this option for our study design, as the consulting dermatologists pointed out that it limited their ability to assess the regions outside of the polygons and, in our test set, all salient regions were contained inside the polygons. This was also indicated in the survey for clarity. However, our polygon approach is, as is, unsuitable for application in the clinic, as different behaviour must be anticipated, especially in degenerate cases. Nevertheless, we believe that a medical device using heatmap-based explanations should offer several interactive modes anyway. We imagine such a tool to allow us to switch between heatmap and polygon explanations and to allow the user to select different levels of opacity or importance thresholds, allowing the user to intuitively determine which reasons are important while limiting interference with the visibility of the lesion. Using a similar scheme as presented in Lucieri et al.[40], we also provide a textual explanation of the characteristics detected in a lesion.

Additionally, in clinical applications, it is essential that the AI system be able to communicate when its predictions are uncertain[41,42]. This allows clinicians to judge when they should trust the AI predictions and when to disregard them. In our study, we communicated the confidence with which the AI used different categorical markers for every predicted characteristic. Predictions of characteristics with high confidence were displayed with the text "strong evidence of X" while those with low confidence were displayed with the text "some evidence of X". We say that the classifier is certain when it finds strong evidence of at least one characteristic (temperature-scaled output above 0.7) and is uncertain otherwise.

We also presented the degree of confidence for the detection of each feature. For classes with temperature-scaled outputs of more than 0.7, we state the predictions as "strong evidence of *feature(s)*" and the other detected features as "some evidence of *feature(s)*" An example of this is provided in Fig. 2a. This communicates the prediction uncertainty to the dermatologist, as the absence of "strong evidence" of all characteristics indicates that the classifier is not confident. This explanation scheme was illustrated to the study participants in a tutorial video in phase 3[2].

The localised explanations in phase 3 were created by showing the regions of interest for the characteristics the classifier was certain about (respective logits above 0.7). If the classifier identified no characteristic above the certainty threshold, we showed the regions of interest for the most certain characteristic instead.

---

[2] https://youtu.be/eWAcaIzXChY

# Study design

Our reader study consisted of three parts and took place between July and December 2022.

## Participants

We recruited a total of 120 international clinicians specialised in dermatology for phase 1, 116 of whom finished the complete study. The participants were contacted via Email through our collaboration network and by using public contact data from the International Society for Dermoscopy website and from university clinic webpages. We also contacted participants from private clinics. As compensation, we offered them the option to be listed as a PubMed indexed collaborator on our work.

We excluded the data of participants who entered constant values for trust, confidence, and/or diagnosis, such as entering a trust score of 7 for all 15 images or a diagnosis of nevus for all 15 images. We excluded images where the participant took less than 7 seconds to complete. We did not use an upper limit for time taken for exclusion because some complicated cases could take a long time to annotate. Furthermore, the participants could pause and resume working later, so a longer time taken did not necessarily imply insincere work. Images marked as having insufficient image quality (n=26) were removed for the particular participant who indicated the issue, but not for others since the image quality issues could have been related to monitor settings. As a result, we evaluated different numbers of images for each participant. None of the participants met these criteria for exclusion. Participants who dropped out in phases 2 or 3 were excluded from the study.

## Phase 1

Phase 1 of the study took place between July and October 2022. We tasked the clinicians to diagnose 15 lesions from our dataset, to explain their diagnoses by choosing the relevant characteristics from our explanatory ontology and to annotate the characteristics in the given images. Furthermore, we asked the clinicians to indicate their confidence in their diagnosis. The participants were informed that they would be presented with 15 lesions (14 unique and one repeated image) each and that this phase would take up to 30 minutes to complete based on the experience from our pilot study. The participants were not informed about the repeated image. The participants were asked to complete this task within two weeks.

We randomly divided the participants into 14 groups. Each group contained roughly 4-6 participants. For each group, we randomly selected 14 images (7 melanomas and 7 nevi) from our test set (196 images in total, with 98 melanomas and 98 nevi) and repeated the third image in the group (either a melanoma or a nevus) after the 12th image. The image sets for each group were mutually exclusive and consisted of 196 unique images and 14 repeated images (see Supplementary Information F for the image IDs). The test set was drawn at random from our dataset and curated to contain only one image per lesion and one lesion per patient. All images from the patients contained in the test set were removed from the training and validation set. We used the repeated image to measure the variability of results for the same participant, but we did not exclude any participants from our analysis based on this variability.

We asked the clinicians to diagnose each lesion as a nevus or melanoma. To reflect the clinical practice of excising lesions that are not considered to be unequivocally benign and the German dermatology guideline to excise specific types of nevi, we offered the diagnoses "nevus (leave in)", "nevus (excise)," and "melanoma". For the evaluation of clinician accuracy, we treated both options for nevus as a simple "nevus" diagnosis.

In addition to the diagnosis, we asked the participants to choose one or more characteristics from the explanatory ontology and to annotate the corresponding image regions (ROIs). Finally, the clinicians were asked to indicate their confidence in their diagnosis on a Likert scale (1-10, with 1 being least and 10 most confident). The participants had the option to indicate issues during the processing of the survey (i.e., "insufficient image quality", "no image visible", "no AI diagnosis visible", and "other", the latter being accompanied by a free text field).

To conduct this part of the study, we used the web-based annotation tool PlainSight[3]. The clinicians received textual information on the study as well as a video[4] explaining use of the tool, the explanatory ontology, and annotation of the ROIs with their login details.

3 participants dropped out in this phase leaving 113 participants.

## Phase 2

The second phase of our study was conducted in November 2022. In this phase, we included the 113 participants who completed phase 1 and 7 participants from our pilot study. We asked the participants who completed phase 1 to diagnose the same lesions they reviewed in phase 1 with the support of an AI system but did not explicitly inform them that they had diagnosed the same lesions in the previous phase. The 7 pilot study participants had not previously reviewed these lesions as we used a different set of images. We ensured that at least two weeks had passed between finishing phase 1 and starting phase 2. Again, the participants were asked to complete the task within two weeks.

The participants were shown the images from phase 1 in the same order alongside the AI diagnosis of the lesion ( "nevus" or "melanoma") and asked them to provide their own diagnosis. As in phase 1, they could choose between "nevus (leave in)", "nevus (excise)" and "melanoma". The participants were informed of the AI's sensitivity and specificity.

As in phase 1, we asked the participants to indicate their confidence in their decisions on a Likert scale (1-10, with 1 being least and 10 being most confident). Additionally, we asked them to indicate their trust in the AI decision on a Likert scale (1-10, with 1 meaning no trust and 10 meaning complete trust in the AI) in this phase.

They were informed that the assessment would take 10-12 minutes to complete. We used the web-based survey tool LimeSurvey[5] to conduct this phase.

3 participants did not complete this phase before the deadline, resulting in 117 participants.

---

[3] https://plainsight.ai/
[4] https://youtu.be/BJRq4nXZ1Xw
[5] https://www.limesurvey.org/

### Phase 3

The final phase of the study was conducted in December 2022. Again, we ensured that at least two weeks had passed between the completion of phase 2 and the start of phase 3. In line with previous phases, the time given for the task was two weeks.

Of the clinicians who completed phase 2, those who participated in phase 3 (n=117) were asked to diagnose the same lesions as in the previous study phases, this time with the support of an explainable AI. Again, they were not informed that they had diagnosed the same lesions in the previous phase or that an image had been repeated, but similar to phase 2, they were informed of the AI's sensitivity and specificity.

For each feature that was detected with certainty (temperature-scaled softmax output > 0.7), we showed a separate explanation. If the AI did not detect any feature with certainty, we showed the explanation for the feature with the highest AI confidence. The participants were informed that they were receiving explanations for "strong evidence for feature(s)" or "weak evidence for feature(s)". The explanations always followed the same schema: the clinicians were shown the relevant entry from the ontology as a textual explanation and the location of the feature based on the highest-influence region(s) of the AI's Grad-CAM saliency map (0.7 or higher). An example is shown in Fig. 2a, b.

Similar to phase 2, the participants were asked to indicate their confidence in their decisions and their trust in the AI decisions. They had the same diagnosis options and the possibility of indicating issues that arose during the assessment. As in phase 2, we used LimeSurvey to conduct this phase and provided the clinicians with a video[6] on how to interpret the AI explanations. A total of 116 participants completed this phase (82 board-certified and 33 resident dermatologists as well as one nurse consultant specialised in dermoscopic skin cancer screening).

### Statistical analysis

All pairwise significance testing was performed using the two-sided paired t test. We utilised the Mann–Whitney U test to determine the significance of the difference between the trust and confidence scores for high- and low-confidence AI predictions and the Wilcoxon signed-rank test to determine the significance of the difference in mean pixel activation ratios due to the nonnormality of the distributions. To calculate confidence intervals, we utilised the bootstrapping method with 10000 samples and a random seed of 42 each time the confidence interval was calculated. We set an alpha value of 0.05, and the P values were adjusted using the Bonferroni method to correct for multiple comparisons.

---

[6] https://youtu.be/eWAcaIzXChY

# Extended Data

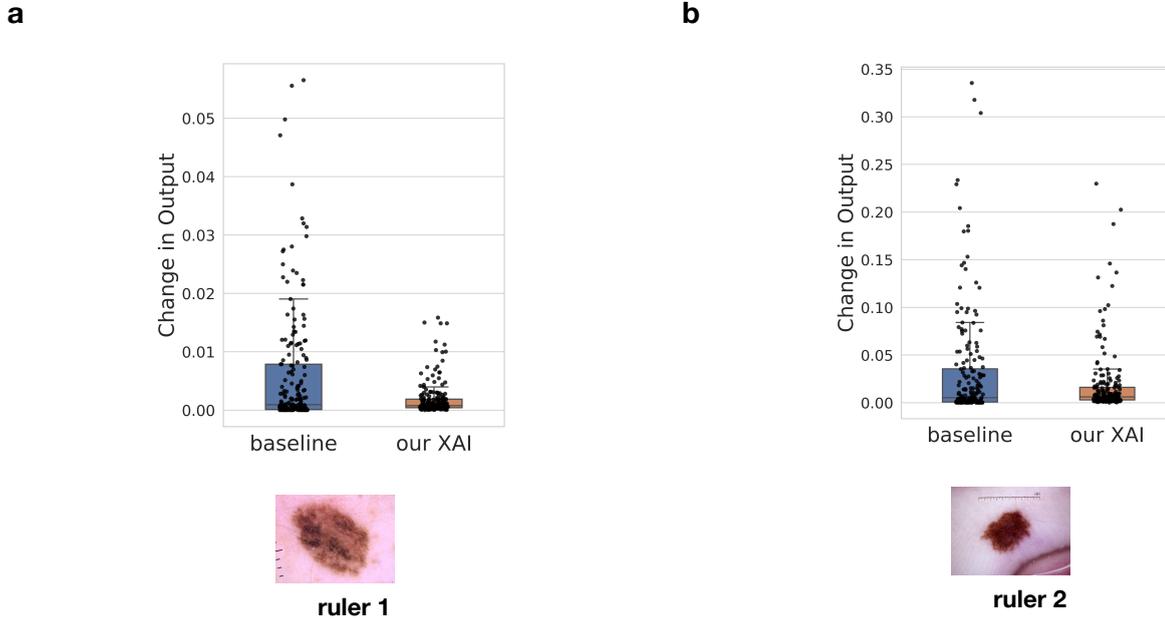

**Extended Data Fig. 1: Robustness of our XAI to artefacts.**

**a**, **b**: Example images of two lesions with superimposed ruler artefacts characterised by dark lines on the images. **c, d:** Sensitivity of our XAI and the baseline classifier to the presence of different rulers. We selected two frequently occurring rulers from our training set and superimposed them on each image in the test set. The y-axis represents the absolute change in classifier logit output after superimposing each of two rulers. Since our XAI has multiple output scores, that is, one for each characteristic, we averaged the output scores per lesion to obtain a single value to facilitate a comparison with the baseline. To assess the sensitivity to artefacts, we computed the mean change in the output score of our XAI and of the baseline classifier when superimposing a ruler on the images. For the images on which we superimposed ruler 1, the mean absolute change in the output score was 0.001 (95% CI: [0.001, 0.002]) for our XAI and 0.006 (95% CI: [0.005, 0.008]) for the baseline classifier (P=P<0.0001, two-sided paired t test, n=196 images). For ruler 2, the mean absolute change in the output score for our XAI was 0.018 (95% CI: [0.014, 0.024]), and that for the baseline classifier 0.032 (95% CI: [0.024, 0.041]) (P=0.005, two-sided paired t test, n=196 images) The central line on each box denotes the median value. The upper and lower box limits denote the 1st and 3rd quartiles, respectively, and the whiskers extend from the box to 1.5 times the interquartile range.

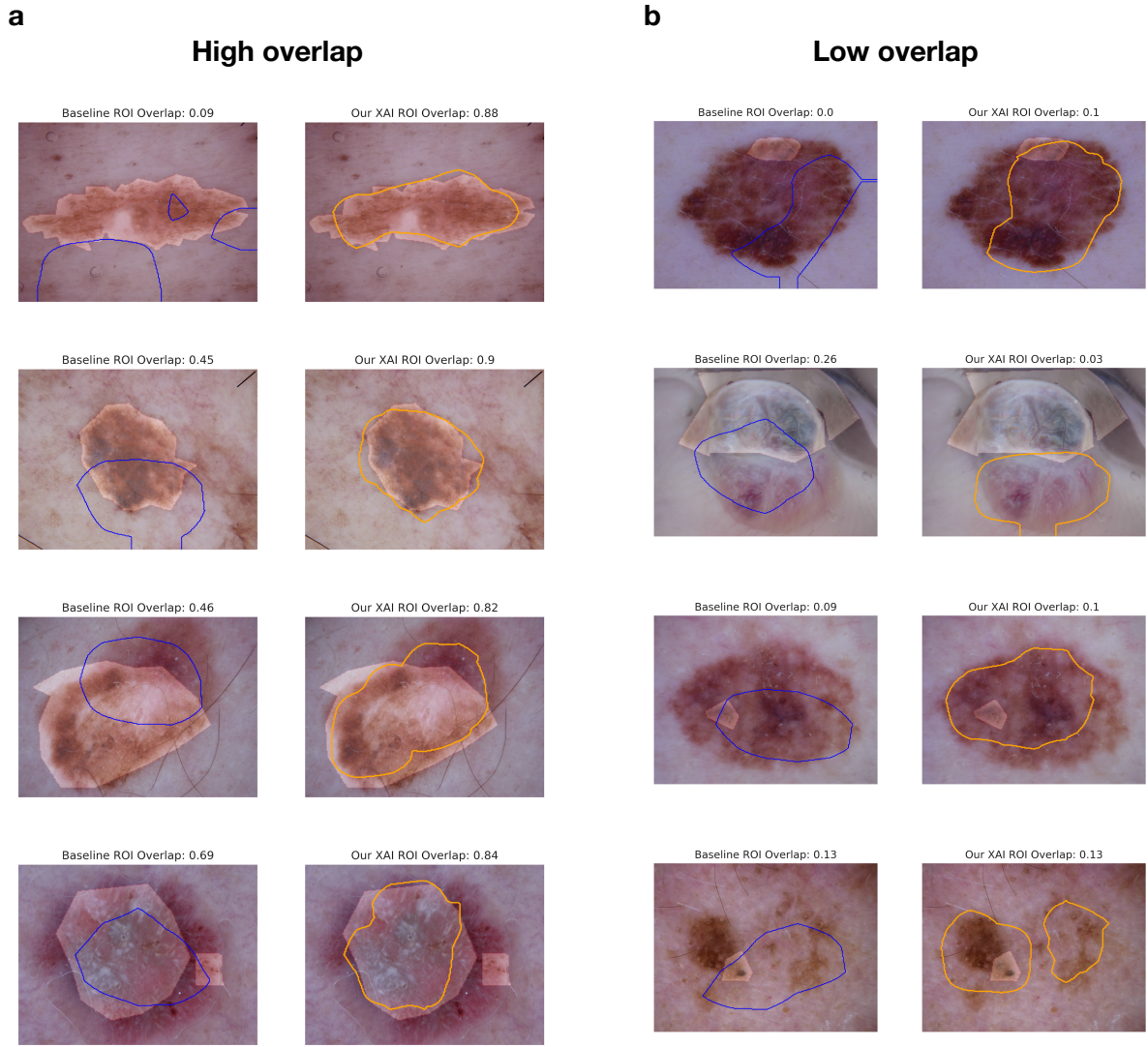

**Extended Data Fig. 2: Examples of high and low overlap between clinician and AI regions of interest (ROIs).**

**a, b**: Examples of low (**a**) and high (**b**) overlap between clinicians and the AI. The ROIs of the baseline classifier are marked with blue polygons, and the ROIs of our XAI are marked with orange polygons.

|  | Sensitivity | Specificity |
|---|---|---|
| **Phase 1** | 59.55% | 72.84% |
| **Phase 2** | 63.86% | 80.75% |
| **Phase 3** | 67.86% | 78.58% |
|  | **Sensitivity (excision)** | **Specificity (excision)** |
| **Phase 1** | 74.46% | 57.64% |
| **Phase 2** | 90.34% | 48.82% |
| **Phase 3** | 90.46% | 45.97% |

**Extended Data Tab. 1: Clinician diagnostic performance in all three phases.**

The top three rows diagnostic sensitivity and specificity when considering a "nevus excise" diagnosis as a nevus (a melanoma that is wrongly diagnosed as a nevus but treated by excision is considered an incorrect choice). The bottom three rows show excision performance (a melanoma that is wrongly diagnosed as a nevus but treated by excision is considered a correct choice).

|  | Confidence | Trust |
|---|---|---|
| **Phase 1** | 6.66 (95% CI 6.43, 6.89) | - |
| **Phase 2** | 7.52 (95% CI 7.43, 7.6) | 6.72 (95% CI 6.61, 6.83) |
| **Phase 3** | 7.78 (95% CI 7.71, 7.86) | 6.96 (95% CI 6.84, 7.1) |

**Extended Data Tab. 2: Mean confidence and trust at the clinician level in all phases.**

Confidence and trust scores were measured on a Likert scale from 1-10, with 1 indicating the lowest confidence/trust and 10 the highest.

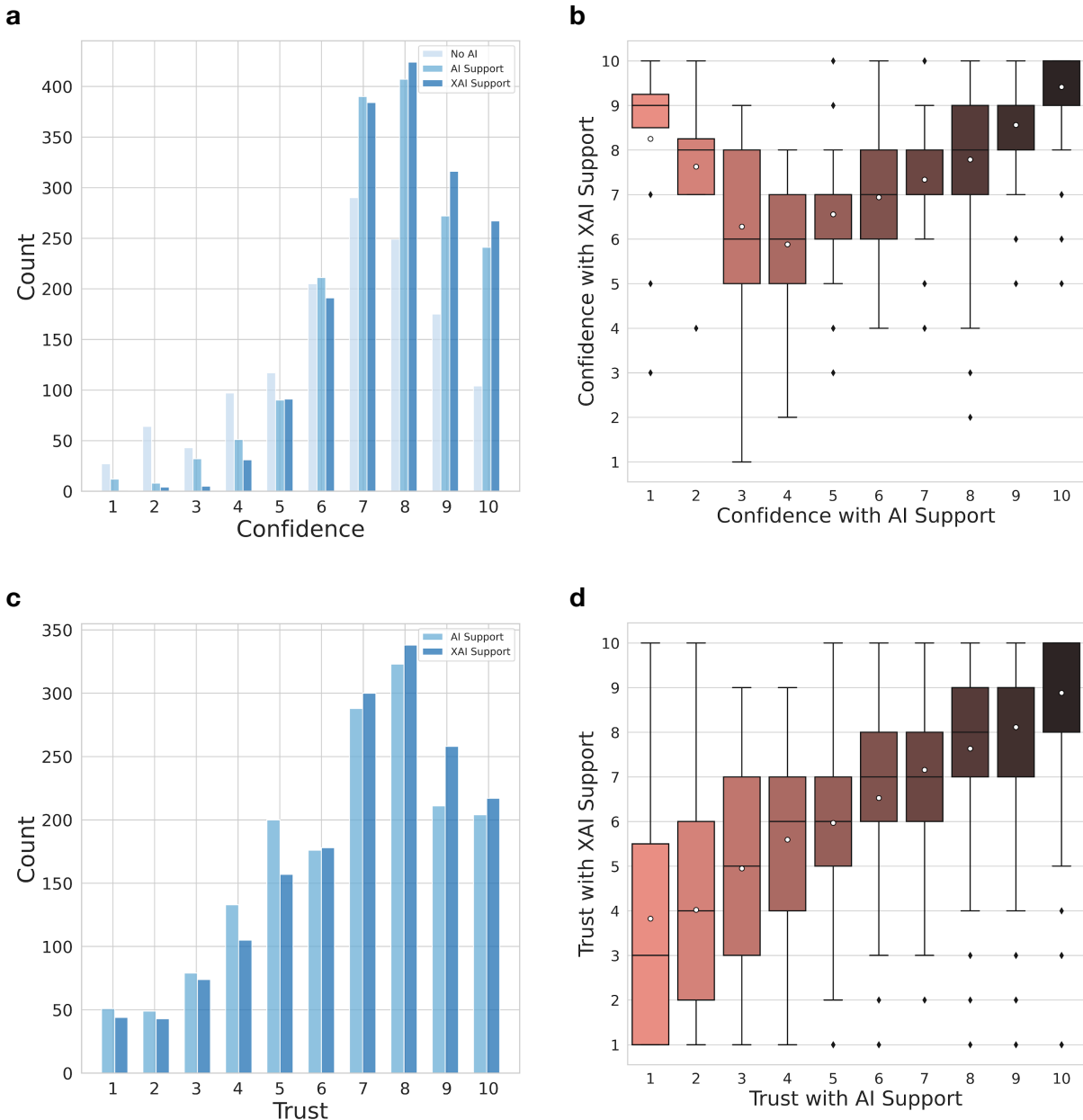

**Extended Data Fig. 3: Confidence and trust scores in all phases.**

**a, c**: Histograms of the confidence (**a**) and trust (**c**) values entered by the clinicians for all phases (phase 1: No AI, phase 2: AI support, phase 3: XAI support). Both confidence and trust were measured on a Likert scale from 1-10, with 1 indicating the lowest and 10 the highest confidence/trust. **b, d**: Box plots of the confidence (**b**) and trust (**d**) values entered by the clinicians for phase 2 (AI support) and phase 3 (XAI support). The figures depict a shift towards higher confidence and trust values as the level of AI support increased.